\documentclass[aip,pop,showpacs,reprint,twocolumn,numerical,superscriptaddress]{revtex4}

\usepackage{graphicx, amsmath}

\begin{document}

\title{A Lagrangian model for the evolution of turbulent magnetic and passive
  scalar fields}
  
\author{T. Hater}
\affiliation{Theoretische Physik I, Ruhr-Universit\"at Bochum,
Universit\"atsstr. 150, D44780 Bochum (Germany)}
\author{H. Homann}
\affiliation{Theoretische Physik I, Ruhr-Universit\"at Bochum,
Universit\"atsstr. 150, D44780 Bochum (Germany)}
\affiliation{Universit\'e de Nice-Sophia Antipolis, CNRS, Observatoire de la
C\^ote d'Azur, Laboratoire Cassiop\'ee, Bd. de l'Observatoire, 06300 Nice,
France}
\author{R. Grauer}
\affiliation{Theoretische Physik I, Ruhr-Universit\"at Bochum,
Universit\"atsstr. 150, D44780 Bochum (Germany)}

\begin{abstract}
In this paper we present an extension of the \emph{Recent Fluid Deformation
(RFD)} closure introduced by Chevillard and Meneveau
\cite{chevillard-meneveau:2006} which was developed for modeling the time
evolution of Lagrangian fluctuations in incompressible Navier-Stokes turbulence.
We apply the RFD closure to study the evolution of magnetic and passive scalar
fluctuations. This comparison is especially interesting since the stretching term
for the magnetic field and for the gradient of the passive scalar are similar but
differ by a sign such that the effect of stretching and compression by the
turbulent velocity field is reversed. Probability density functions (PDFs) of
magnetic fluctuations and fluctuations of the gradient of the passive scalar
obtained from the RFD closure are compared against PDFs obtained from direct
numerical simulations.
\end{abstract}



\maketitle
  
\section{Introduction}
In recent years several Lagrangian type closures have been developed to model
fluctuations of turbulent incompressible fields
\cite{chertkov-pumir-etal:1999,naso-pumir:2005,chevillard-meneveau:2006,chevillard-meneveau:2007}.
The starting point for all these models goes back to the so-called
\emph{Restricted Euler} closure \cite{viellefosse:1984,cantwell:1992} which
models the time evolution of the gradient tensor
$\mathcal{A}_{ij} := \partial_i u_j$
\begin{equation}
	\label{eq:re}
	\frac{\partial}{\partial t}\mathcal{A}_{ij} 
	= - \left( \mathcal{A}_{ik} \mathcal{A}_{kj} - \frac{\delta_{ij}}{3}
	\mathcal{A}_{mk} \mathcal{A}_{km} \right)\;\; .
\end{equation}
This model is derived using the incompressibility constraint and neglects the
viscous term and the anisotropic part of the pressure Hessian. The
\emph{Restricted Euler} model is already able to capture geometric features of
vortex stretching and alignment but fails as a robust model to obtain stationary
statistics due to the appearance of finite time singularities.

Several strategies have been applied to regularize the restricted Euler model
where the most prominent ones are the \emph{Tetrad Model}
\cite{chertkov-pumir-etal:1999} which models the anisotropic part of the
pressure Hessian and the \emph{Recent Fluid Deformation} approach which models
both the viscous term and the anisotropic part of the pressure Hessian
\cite{chevillard-meneveau:2006}.

In this paper we extend the \emph{Recent Fluid Deformation} approach to the case
of the time evolution of the gradient of passive scalar and to kinematic
magnetohydrodynamic turbulence. Comparing these two examples is especially
interesting since both examples differ mainly by a sign in the stretching term.
Thus structures which are expanded in one equation (e.g. gradient of passive
scalar) are compressed in the other (e.g. MHD). Comparison with direct
numerical simulations are performed to test the range of applicability of the
extended \emph{Recent Fluid Deformation} models.

\section{Velocity gradients}

Before we apply the \emph{Recent Fluid Deformation} approach to the gradient of
a passive scalar and to kinematic MHD turbulence, we first recall the main steps in
the derivation of this model. Detailed information can be found in
\cite{chevillard-meneveau:2006,chevillard-meneveau:2007}. Starting with the
Navier-Stokes equation 
\begin{equation}
	\label{eq:ns}
	\partial_t \mathbf{u} + \mathbf{u}\cdot\nabla\mathbf{u} + \nabla p = \nu
	\Delta \mathbf{u} \;\; , \;\;\; \nabla\cdot\mathbf{u} = 0
\end{equation}
one takes the gradient
\begin{equation}
	\label{eq:gradns}
    \partial_t\partial_k u_i + \partial_k (u_j\partial_ju_i)
    + \partial_{ki}p = \nu \partial_{kjj}u_i
\end{equation}
and denotes the velocity gradient $\partial_iu_j$ as $\mathcal{A}_{ij}$. Using
the Lagrangian material derivative $\frac{\mathrm{d}}{\mathrm{d}t} =
(\partial_t + u_j\partial_j)$ eqn. (\ref{eq:gradns}) reduces to
\begin{equation}
    \label{eq:sde_ns}
	\frac{\mathrm{d}\mathcal{A}_{ij}}{\mathrm{d}t}
    +\mathcal{A}_{ik}\mathcal{A}_{kj} = -\partial_{ij}p
    +\nu \partial_{kk}\mathcal{A}_{ij}\,\,\mathrm{,}
\end{equation}
Assuming the pressure Hessian to be isotropic $\partial_{ij} p =
\frac{\delta_{ij}}{3} \partial_{kk} p$ and neglecting the viscous term would
result in the Restricted Euler model given by eqn. (\ref{eq:re}). In order to
improve the model for the pressure Hessian, Chevillard and Meneveau
\cite{chevillard-meneveau:2006} consider the pressure Hessian $P_{nm}=
\frac{\partial^2p}{\partial X_n\partial X_m}$ expressed in the Lagrangian frame
and postulate that the Lagrangian pressure is isotropic at some given time $t_0$.
Applying the change between Eulerian and Lagrangian coordinates 
$\frac{\partial}{\partial x_i} = \frac{\partial X_j}{\partial
x_i}\frac{\partial}{\partial X_j}$ twice - while neglecting higher orders - one
obtains
\begin{equation}
    \frac{\partial^2 p}{\partial x_i\partial x_j} \simeq\frac{\partial X_n}{\partial x_i}\frac{\partial X_m}{\partial x_j}\frac{\partial^2 p}{\partial X_n\partial X_m}
    = -\frac{\mathcal{A}_{kl}\mathcal{A}_{lk}}{C^{-1}_{kk}}C^{-1}_{ij} 
\end{equation}
where $\mathbf{C} =\exp(\tau \boldsymbol{\mathcal{A}})\exp(\tau
\boldsymbol{\mathcal{A}}^T)$ is the \textsl{short-time Cauchy-Green tensor}
resulting from the integration of the flow map. The time $\tau$ is chosen of
the order of the Kolmogorov time so that it is possible to linearize the
integration from $\partial_i u_j$ to $\partial_i X_j$.

Applying a similar reasoning to the viscous term results in the approximation
$\nu\Delta\mathcal{A}_{ij} \simeq -\frac{C^{-1}_{kk}}{3T}\mathcal{A}_{ij}$, $T$
being a characteristic friction time such that the Reynolds number $\mathcal{R}$
is proportional to $\left(\frac{T}{\tau}\right)^2$
\cite{chevillard-meneveau:2006}.
In all simulations described below, the parameters $T=1$ and $\tau=0.05$ were
chosen which result in a Taylor-Reynolds number of $\mathcal{R} = 77$. This value is in the
range of the performed DNS simulations (see Table \ref{table:paramscalar}).

Finally, one obtains a stochastic ordinary differential equation (SDE)
\begin{equation}
    \label{eq:final}
    \mathrm{d}\mathcal{A}_{ij} = -\left(\mathcal{A}_{ik}\mathcal{A}_{kj} 
    -\frac{\mathcal{A}_{kl}\mathcal{A}_{lk}}{C^{-1}_{kk}}C^{-1}_{ij}
    +\frac{C^{-1}_{kk}}{3T}\mathcal{A}_{ij} \right)\mathrm{d}t +
    \mathrm{d}W_{ij}
\end{equation}
where the stochastic forcing term $\mathbf{W}$ is added to represent large scale
forcing.
The SDE (\ref{eq:final}) can be integrated with standard methods for
stochastic equations \cite{kloeden-platen:1999,higham:2001}.

\begin{figure}[t!]
	\centering
	\includegraphics[width=0.4\textwidth,angle=-90]{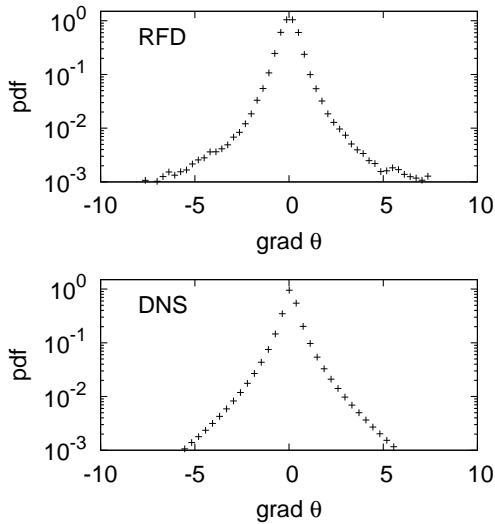}
	\caption{The PDF of the gradient of the passive scalar
	$\nabla \theta$ using the \emph{Recent Fluid Deformation} approach (top) and
	obtained from direct numerical simulations (bottom).
	\label{fig:gradtheta}}
\end{figure}

\section{Extension to passive admixtures}

In this section we apply the \emph{Recent Fluid Deformation} theory to the case
of an admixture $\theta$ which is passively advected by the fluid according to
\begin{equation}
    \label{eq:ps}
    \partial_t\theta + \mathbf{u}\cdot\nabla\theta = \kappa\Delta\theta \,\,.
\end{equation}
In order to study the dynamics of the gradient of the passive field $\theta$ 
we take the derivative of eqn.~(\ref{eq:ps}) and obtain
\begin{equation}
    \label{eq:gradps}
    \partial_t (\partial_i\theta) + \partial_i
    (u_k\partial_k\theta) = \kappa\partial_{kk} (\partial_i
    \theta) \,\,.
\end{equation}
Using again the Lagrangian material derivative $\frac{\mathrm{d}}{\mathrm{d}t} =
(\partial_t + u_j\partial_j)$ eqn. (\ref{eq:gradps}) reduces to
\begin{equation}
	\label{eq:gradps2}
	\frac{d}{dt} (\nabla \theta)_i = - \mathcal{A}_{ik} (\nabla \theta)_k 
	+ \kappa\partial_{kk} (\nabla \theta)_i \,\,.
\end{equation}
In this equation, the gradient tensor $\mathcal{A}_{ij}$ is obtained from
integrating eqn. (\ref{eq:final}). Modeling the viscous term as for eqn.
(\ref{eq:sde_ns}) and adding a stochastic forcing $\mathbf{V}$, we obtain the
stochastic differential equation (SDE)
\begin{equation}
    \label{eq:sde_ps}
    \mathrm{d}(\nabla\theta)_i = -\left[\mathcal{A}_{ik}(\nabla\theta)_k
    +\frac{C^{-1}_{kk}}{3T_\theta}(\nabla\theta)_i\right]\mathrm{d}t +
    \mathrm{d}V_i
\end{equation}
This equation is integrated together with the equation (\ref{eq:final}) for the
velocity gradient tensor. The forcing is chosen to be vectorial 
gaussian noise, delta-correlated in time, as for the velocity gradient.
The forcing amplitude is $\sqrt{2\mathrm{d}t}$ as for a standard random walk.
The parameters for the result shown here are $T_\theta= 1$ and $\tau=0.05$.

In order to test the applicability of the \emph{Recent Fluid Deformation}
approximation, we performed direct numerical simulations for passive scalar
turbulence. For this, we utilized the pseudo-spectral simulation framework
\textsc{LaTu} \cite{homann-kamps-etal:2009}. Table \ref{table:paramscalar}
summarizes the relevant parameters of the simulation.

This model extension has been analyzed in \cite{gonzalez:2009} in
great detail, considering important phenomena like amplification of the gradient's norm 
and alignment to strain principal axes. But in contrast to this publication, we focus 
on the comparison of different turbulent systems.

\squeezetable
\begin{table*}
  \centering
  \begin{ruledtabular}
 \begin{tabular}{ccccccccc}
                   & $N^3$   & $\mathcal{R}_{\lambda}$& $u_\mathrm{rms}$ & $\epsilon_\mathrm{k}$ & $\nu$            & dx       & $\eta$              & L      \\
   
   passive scalar  & $256^3$ & $106$                  & $0.0321$         & $0.0015$              & $4\times 10^{-4}$ & $0.0245$ & $1.43\times 10^{-2}$ & $2.02$  \\
   
   MHD	           & $256^3$ & $68$                   & $0.41$           & $0.047$               & $2\times 10^{-3}$ & $0.0245$  & $0.02$             & $1.47$
 \end{tabular}
 
  \end{ruledtabular}
 \caption{Parameters of the numerical simulations.
    $N^3$: number of collocation points, 
    $\Re_\lambda = \sqrt{15 u_\mathrm{rms} L/ \nu}$: Taylor-Reynolds number,
    $u_\mathrm{rms}$: root-mean-square velocity,
    $\epsilon_\mathrm{k}$: mean kinetic energy dissipation rate, 
    $\nu$: kinematic viscosity,
    $dx$: grid-spacing,
    $\eta =(\nu^3/\epsilon_\mathrm{k})^{1/4}$: Kolmogorov dissipation length scale,
    $L = (2/3E)^{3/2}/\epsilon_\mathrm{k}$: integral scale.
    Schmidt number and magnetic Prandtl number are equal to 1.}
  \label{table:paramscalar}
\end{table*}

The top of Figure \ref{fig:gradtheta} shows the probability distribution
function (PDF) of fluctuations of $\nabla \theta$ obtained from integrating eqns.
(\ref{eq:final}) and (\ref{eq:sde_ps}). In the bottom of
Figure~\ref{fig:gradtheta} the PDF obtained from the direct numerical
spectral simulation is shown. The qualitative agreement
is excellent and the tails of the PDFs of $\nabla \theta$ approximately follow
an exponential decay behavior as predicted by Shraiman and Siggia
\cite{shraimann-siggia:1994} for a Gaussian velocity field.

\begin{figure}[t!]
	\centering
	\includegraphics[width=0.4\textwidth,angle=-90]{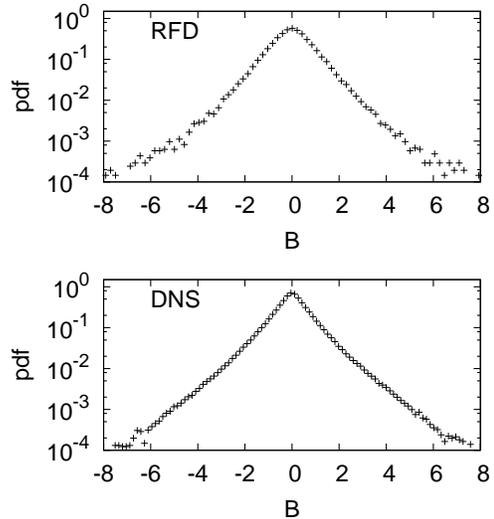}
	\caption{The PDF of the magnetic field $B$
	using the \emph{Recent Fluid Deformation} approach (top) and
	obtained from direct numerical simulations (bottom).
	\label{fig:B}}
\end{figure}

\section{Extension to the MHD equations}

We start with the incompressible MHD equations in the kinematic regime
neglecting the back-reaction of magnetic fluctuations on the velocity
field. The equation for the time evolution of the magnetic field
$\mathbf{B}$ reads
\begin{equation}
	\label{eq:mhd}
   \partial_t \mathbf{B} + \mathbf{u}\cdot\nabla \mathbf{B} =
   \mathbf{B} \cdot \nabla \mathbf{u} + \eta \Delta \mathbf{B} \;\; ,
\end{equation}
where the velocity field obeys the Navier-Stokes equations~(\ref{eq:ns}).
In the Lagrangian frame this equation takes the form
\begin{equation}
	\label{eq:mhdlag}
   \frac{d}{dt} B_i = B_k A_{ki} + \eta \Delta \mathbf{B} \;\; .
\end{equation}
Again, using the same \emph{Recent Fluid Deformation}
approximation for the resistive term and adding a stochastic forcing, which is as above gaussian 
and white-in-time, we obtain
\begin{equation}
   \label{eq:sde_mhd}
    \mathrm{d}B_i = -\left(-B_k\mathcal{A}_{ki} 
                  + \frac{C^{-1}_{kk}}{3T_B} B_i \right) dt + \mathrm{d}U_i
                  \,\,.
\end{equation}

Note that this equation differs from \eqref{eq:sde_ps} by the transpose in
$\boldsymbol{\mathcal{A}}$ and - more important - the sign of the first term on
the right hand side. Thus directions, where the magnetic field is stretched, are
directions, where the gradient of the passive scalar is compressed and vice
versa. This very different influence of the stretching term results in a
qualitatively different PDF for the magnetic field fluctuations compared to the
$\nabla \theta$ fluctuations. In order to test whether the \emph{Recent Fluid
Deformation} approach is able to reproduce the different role of the stretching
term, we performed direct numerical simulations of kinematic MHD turbulence.
Here, we again performed simulations using the framework \textsc{LaTu} and the
relevant parameters are given in Table \ref{table:paramscalar}. The simulation
was started with random initial condition for the magnetic field $\mathbf{B}$
and an already fully turbulent velocity field. External forcing was only applied to the
velocity field by keeping the large scale Fourier-modes $k\le 2$ constant.
The top of Figure \ref{fig:B} shows the PDF of magnetic field fluctuations
calculated using equations (\ref{eq:final}) and (\ref{eq:sde_mhd}). This PDF has
to be compared to the PDF obtained from direct numerical simulation shown in the
bottom of Figure \ref{fig:B}. It is again remarkable that the simple stochastic
ODE model (\ref{eq:sde_mhd}) is able to capture correctly the shape of the PDF
of magnetic field fluctuations.

Since we considered the kinematic MHD equations in the Dynamo regime, the PDF of
the stochastic model could not be obtained from a time series as in the case of
the passive scalar since the magnetic field is growing exponentially. 

To obtain meaningful statistics ensemble averaging was used instead of time sampling.
Numerically, a random initial state was generated and integrated a sufficiently large 
number of time steps, to retain no statistical influence of the initial state. 
The final state enters the statistics and a new realization is generated via another random
initial condition. The PDF was obtained by sampling over $10^6$ initial
conditions.

\begin{figure}[b!]
	\centering
	\includegraphics[width=0.33\textwidth,angle=-90]{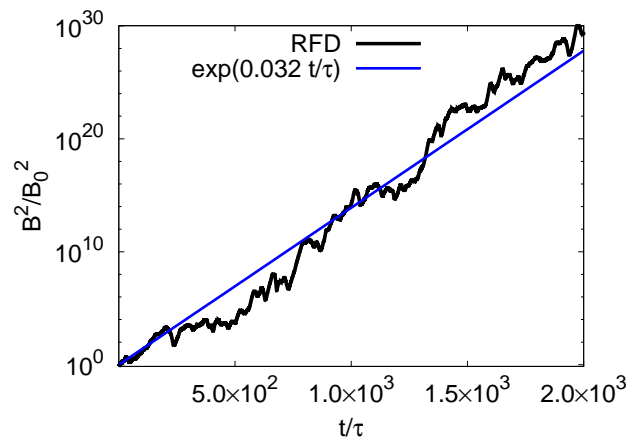}
	\caption{(Color online) Temporal growth of the magnetic field energy. To be well in the dynamo regime  we chose $T_B = 6$.
	\label{fig:dynamo}}
\end{figure}

For illustration of the kinematic dynamo effect and a consistency check with
direct numerical simulations of the kinematic MHD equations we tracked the local
magnetic energy $\mathbf{B}^2(\mathbf{X}(t),t)$ in Figure \ref{fig:dynamo}
for $T_B=6$, a value well in the dynamo regime of the modell.
For this simulation only the velocity gradient itself was
driven by stochastic forcing, as outlined above, while the magnetic field was
undriven. The forcing amplitude was  scaled with $\sqrt{2\mathrm{d}t}$, analog
to the scaling for a standard random walk. \\
The parameter $\tau$ of the model was chosen to be $\tau = 0.05$. \\
The magnetic diffusion time $T_B$
scale is varied to explore the reaction of the dynamo effect. Below the critical
parameter of approximately $T_B\approx 4$ we observed no growing magnetic
energy.

The corresponding growth rate, estimated from an exponential growth $\approx
\exp(\gamma t/\tau)$, is about $\gamma \simeq 0.032$. The growth rate compares
well with the growth rate of our DNS simulation, where we obtain $\gamma \simeq
0.03$.  Although the identification of the correct Kolmogorov timescale $\tau$
is not without uncertainties, it is remarkable that this value of the growth
rate is in the range of data obtained by kinematic dynamo simulations (see
also Table I in \cite{eyink:2010} and references therein).

\section{Conclusion \& outlook}
In this paper we have shown that the natural extension of the
\emph{Recent Fluid Deformation} model for Navier-Stokes turbulence
\cite{chevillard-meneveau:2006} to the case of fluctuations of the gradient of a
passive scalar and to magnetic field fluctuations is able to produce probability
distribution functions that agree well with PDFs obtained from direct numerical
simulations. The PDFs from the stochastic ODEs (\ref{eq:final}, \ref{eq:sde_ps},
\ref{eq:sde_mhd}) are obtained with a fraction of the computing resources
necessary for direct numerical simulations. The next step is to include the
back-reaction of the magnetic field on the fluid flow. If this can be managed,
then generating magnetic field fluctuations using the \emph{Recent Fluid
Deformation} model is a tempting alternative for e.g. the problem of cosmic ray
propagation where magnetic field fluctuations are generated by other means which
are not able to capture intermittency effects (see
\cite{giacalone-jokipii:1999,zimbardo-pommois-veltri:2006}). \\

\acknowledgements
R.G. acknowledges stimulating discussions with L. Chevillard during the
workshop ``Euler Equations: 250 Years On''.
Access to the JUGENE BlueGene/P computer at the FZ J\"ulich was made available
through project HBO22.
This work benefited from support through DFG-FOR1048.

\end{document}